%% file: paper.tex
\newif\ifarxiv\arxivtrue
\newcommand\ARXIV[1]{\ifarxiv #1 \fi}
\newcommand\VMCAI[1]{\ifarxiv \else #1 \fi}

\ifarxiv
\documentclass[runningheads]{llncs}
\else
\documentclass{llncs}
\fi

\usepackage{main/mypackage}

\newcommand{\supplementlink}{https://doi.org/10.5281/zenodo.14170859}

\VMCAI{
  \usepackage[firstpage]{draftwatermark}
  \SetWatermarkAngle{0}
  \SetWatermarkText{\hspace*{5.5in}\raisebox{9in}{\href{\supplementlink}{\includegraphics{badge-afr}}}} 
}

\begin{document}

\title{An Abstract Domain for Heap Commutativity}
\ARXIV{
  \subtitle{(Extended Version)}
}

\author{
  Jared Pincus\inst{1}\orcidID{0000-0001-6708-5262} \and
  Eric Koskinen\inst{2}\orcidID{0000-0001-7363-634X}
}

\authorrunning{J. Pincus and E. Koskinen}

\institute{
  Boston University, Boston MA 02215, USA
  \email{pincus@bu.edu} \and
  Stevens Institute of Technology, Hoboken NJ 07030, USA
  \email{eric.koskinen@stevens.edu}
}

\maketitle

\begin{abstract}
  \input{main/abstract.tex}
\end{abstract}

\input{main/body.tex}

  \begingroup\small
  \subsubsection{Acknowledgments.}
  \input{main/acknowledgements.tex}

  \subsubsection{Disclosure of Interests.}
  \input{main/disclosures.tex}
  \endgroup

\bibliographystyle{splncs04}
\bibliography{main/paper}

\ifarxiv
\input{appendices/isomorphism.tex}
\fi

\end{document}

%% file: main/abstract.tex


Commutativity of program code 
(\emph{i.e.}~the equivalence of two code fragments composed in alternate orders)
is of ongoing interest in many settings such as
program verification, scalable concurrency, 
and security analysis.
While some have explored static analysis for code commutativity,
few have specifically catered to heap-manipulating programs.
We introduce an
abstract domain
in which commutativity synthesis or verification techniques
can safely be performed on abstract mathematical models
and, from those results, one can
directly obtain commutativity conditions for concrete heap programs.
This approach offloads challenges of concrete heap reasoning
into the simpler abstract space.
We show this reasoning supports framing and composition, and conclude with commutativity analysis of programs operating on example heap data structures.
Our work has been mechanized in Coq and is available in the supplement.

%% file: main/keywords.tex
Commutativity \and
Abstract Interpretation \and
Observational Equivalence \and
Separation Logic

%% file: main/body.tex
\input{sections/intro.tex}

\input{sections/overview.tex}

\input{sections/prelim.tex}

\input{sections/define.tex}

\input{sections/semantics.tex}

\input{sections/commutativity.tex}

\input{sections/conjunction.tex}

\input{sections/full-examples.tex}

\input{sections/related-work.tex}

\input{sections/conclusion.tex}

%% file: sections/intro.tex
\section{Introduction}
\label{sec:introduction}

Commutativity describes circumstances in which
two programs executed in sequence yield the same result
regardless of their execution order.
Fundamentally, commutativity offers a kind of independence
that enables parallelization or 
simplifies program analysis by collapsing equivalent cases.
As such, this property has long been of interest in many areas 
including databases~\cite{weihl}, 
parallelizing compilers~\cite{RinardDiniz}, 
scalable~\cite{scalable_commutativity} and
distributed systems~\cite{shapiro2011comprehensive}, 
and smart contracts~\cite{Dickerson2017,sergey_sharding}.
Commutativity is also used in verification,
including efforts to simplify
proofs of parameterized programs~\cite{FarzanPOPL2024}, 
and concurrent program properties~\cite{FarzanPOPL2023}
such as termination~\cite{FarzanCAV2023},
dynamic race detection~\cite{DBLP:journals/pacmpl/FarzanM24}, and information flow~\cite{EilersPLDI2023}.

Efforts in \emph{commutativity analysis} seek to infer or verify
\emph{commutativity conditions}---%
the initial conditions (if any)
under which code fragments commute.
For example, Rinard and Kim~\cite{rinard_linked_structures} verify 
commutativity conditions by 
relating a data structure to its pre/postconditions, then 
showing commutativity of that abstraction. 
Later, methods were developed to automatically synthesize commutativity conditions for specifications and imperative program fragments~\cite{bansal_tacas,JAR20,ATVA23,veracity_paper},
which can be used to automatically parallelize programs~\cite{RinardDiniz,CommSet,veracity_paper}.

\emph{Objectives.}
Considering its many applications, 
it would be appealing 
to extend commutativity analysis to the heap
memory model.
However, few works thus far
have targeted heap-modifying programs.
P\^{i}rlea \emph{et al.}~\cite{sergey_sharding}
describe a commutativity analysis with some heap treatment,
but with the limited scope of heap disjointness/ownership.
Eilers \emph{et al.}~\cite{EilersPLDI2023}
employ, but do not verify,
commutativity to reason about information flow security
in concurrent separation logic. 
This paper thus aims to analyze commutativity of heap programs.

\emph{Challenges \& Contributions.}
Commutativity reasoning can be challenging for heap-manipulating programs,
but is more straightforward 
for functions over simple mathematical objects.
With that in mind,
this paper presents an abstract domain
which is amenable to existing commutativity condition
verification and synthesis techniques~\cite{bansal_tacas,JAR20,ATVA23},
and allows conditions found in said abstract space
to immediately yield concrete separation-logic-style conditions for heap programs.
For instance,
given a simple mathematical list representation of a stack data structure,
we can easily find that the functions \emph{push} and \emph{pop} operating on these lists
commute when the first (i.e.\ top) list element equals the value being pushed.
We may then thread this condition down
to concrete \tt{push} and \tt{pop} methods.
For this threading to be sound,
we design the domain to handle
various complications of heap programs,
including observational equivalence of heap data structures,
nondeterminism of address allocation,
and improperly allocated heaps and resultant program failure.

In summary, the contributions of this paper include:

(Sec.~\ref{sec:construction}) 
We introduce a simple abstract domain
that is amenable to abstract commutativity reasoning,
which can soundly entail concrete commutativity.
The domain is designed to lift
heap-allocated structures to 
mathematical objects (e.g.\ value lists to encode a stack)
via separation-logic-style predicates,
while cleanly handling ambiguous or improperly allocated heaps.
We then describe the semantics of abstract programs,
and establish a soundness relationship
between abstract and concrete programs
in the style of abstract interpretation~\cite{cousot-cousot-abstract-interp},
which enables the reduction of concrete to abstract commutativity.

(Sec.~\ref{sec:commutativity}) 
We define concrete and abstract commutativity,
and present our main soundness theorem:
that a valid commutativity condition for two concrete programs 
may be derived immediately from commutativity reasoning performed (much more easily) on their corresponding abstract programs.
These derived concrete conditions are also \emph{framed}, i.e.\ if the programs commute under a given initial heap layout,
they also commute in any larger layout.

(Sec.~\ref{sec:conjunction}) 
We define a way to compose abstractions, 
which mirrors the separating conjunction on concrete heaps,
and present compositional commutativity properties.
Composition aids in settings with multiple allocated data structures,
and working with arguments/return values.

(Sec.~\ref{sec:full-examples}) We share four complete examples:
a non-negative counter,
two-place unordered set,
linked-list stack,
and the composition of a counter and stack.

\VMCAI{(Supplement)}%
\ARXIV{\href{\supplementlink}{(Supplement)}}%
Our work is implemented in Coq,
on top of Separation Logic Foundations \cite{sep-logic-foundations}.
All theorems, lemmas and properties have been proved in Coq
(except for the example programs written in the concrete semantics of \S\ref{subsec:particular-semantics}).

\emph{Limitations.}
In this paper we establish the theory necessary
for deriving heap commutativity from abstract commutativity reasoning.
We leave implementation of this theory,
including automation and synthesis,
for future work.

%% file: sections/overview.tex

\section{Overview}
\label{sec:overview}

We summarize our work through the example
of a stack, implemented as a heap-allocated linked list,
with typical \tt{push} and \tt{pop} methods
(\tt{pop} fails if the stack is empty).
For simplicity, method arguments or return values are stored in method-specific pre-allocated heap cells (e.g.\ $a$).
  \begin{center}
    \input{conc_progs/stk-push-wide.tex}
    \hspace*{1em}
    \input{conc_progs/stk-pop-wide.tex}
  \end{center}

\emph{Inconvenience of concrete heap commutativity.}
We would like to know under what precondition $P$, if any,
do \tt{push} and \tt{pop} \emph{commute}.
Verifying whether \tt{assume}($P$);\tt{push};\tt{pop} is equivalent to \tt{assume}($P$);\tt{pop};\tt{push} is not amenable to typical
relational reasoning techniques~\cite{Yang2007},
because the swapped orders of \tt{push} and \tt{pop} (or, in general, $f$ and $g$) do not yield meaningful alignment points
\cite{%
NagasamudramN21,AntonopoulosKLNNN23}.
So, the problem amounts to performing four forward symbolic executions
per method pair, meaning quadratically many executions
as the number of methods grows.
There are, however, efficient CEGAR/CEGIS methods for verifying and even synthesizing a condition $P$ of logical method pre/postconditions~\cite{bansal_tacas,JAR20,ATVA23}. We now discuss a simple abstraction of the Stack methods as mathematical functions,
how our theory relates the concrete heap to this abstraction, 
and how this enables \emph{concrete} commutativity conditions to be immediately derived from the results of simpler (automatable) 
abstract commutativity reasoning.

\emph{Abstract commutativity.}
Let us, for the moment, set aside the concrete heap implementation, and consider a simple mathematical representation of a stack as a list of values, along with functions
over the stack with an input/output ``cell'':
\begin{center}
$
    \ap{push}(s,v) \coloneqq (v \cons s, v) 
    \hspace*{1.5em}
    \ap{pop}([],\_) \coloneqq \bot
    \hspace*{1.5em}
    \ap{pop}(v \cons s, \_) \coloneqq (s, v)
$
\end{center}
Here, $\ap{push}$ appends the input value to the stack, 
and $\ap{pop}$ removes and returns the topmost value,
but fails (yields $\bot$) on an empty stack. 
Reasoning about commutativity of these mathematical objects is convenient, and synthesizing commutativity conditions can now be automated~\cite{bansal_tacas,JAR20,ATVA23}.
For this example,
given an initial stack $s$,
$\ap{push}(u)$ and $\ap{push}(v)$ commute when $u=v$;
$\ap{push}(u)$ and $\ap{pop}()$ when $s = u \cons \_$;
and $\ap{pop}()$ and $\ap{pop}()$ when $s = x \cons x \cons \_$ for some $x$.

The question we ask in this paper
is how we might
abstract the concrete programs so that, 
from these abstract commutativity conditions,
we can immediately obtain concrete conditions.
We show that this is indeed possible;
for example, in the case of $\tt{push}_\ell^a$ and $\tt{push}_\ell^b$ we obtain the
(separation logic) precondition:
$\exists u.\ \PsiStack^+(\_) \ast a \mapsto u \ast b \mapsto u$,
where $\PsiStack^+(\_)$ asserts that the heap
contains a properly allocated stack
(we formally define $\PsiStack^+$ below).
We now summarize how to achieve this using the Stack example.

{\bf Step 1: Projecting heaps to abstract Stacks.}
To soundly lift concrete heap structures into an abstract domain,
we must accommodate various complications of the heap,
such as improperly allocated heaps,
observationally equivalent layouts,
and allocation nondeterminism.
We achieve this by
choosing a set of abstract values $X$ 
(such as value lists to represent stacks),
and a projection function $\pi$ which maps
concrete heaps into $X$.
Whether a heap is ``well-structured'' depends on the choice of $\pi$.
For the Stack, $\pi$ 
encodes that the only valid heaps for the abstraction
are those containing a correctly allocated linked stack.
We say that such heaps 
are ``within the purview'' of the abstraction at hand,
meaning they will be amenable to abstract commutativity reasoning.

Of course, $\pi$ must also account for heaps which are \emph{not} properly
allocated to be representable by a value in $X$.
To this end, every abstract domain includes a special element $\opurv$ (``cross''),
to which $\pi$ maps all inappropriately structured, or ``out-of-purview'', heaps.
Commutativity w.r.t.\ such heaps cannot be reasoned about meaningfully in the abstract space.

For our Stack example,
the abstract values $X$ are lists of non-\vnull\ concrete values,
and the projection function $\pi$ maps a heap to such a list
by checking (via separation logic heap predicates~\cite{Parkinson})
whether it contains a valid linked list,
and extracting the contents of said list.
Precisely:
\begin{center}
\input{figures/stk-abstr-overview.tex}
\end{center}
By existentially quantifying the intermediate addresses of the linked stack
and omitting them from the extracted abstract value,
we render stacks observationally equivalent
if and only if their contained values match.

{\bf Step 2: Abstract domain for commutativity.}
The next question is how this projection should induce an abstract domain, 
while soundly respecting concrete programs, and enabling commutativity reasoning.
The key is to ``wrap'' the abstract values $X$
into a domain that also includes
elements which track if abstract commutativity still 
entails concrete commutativity.
Namely, we add %
the abstract value $\opurv$ for out-of-purview heaps,
$\ipurv$ (``check'') for an ambiguous collection of in-purview heaps,
$\bot$ as a failure state,
and $\top$ for a collection of heaps both in- and out-of-purview.
Instrumenting $\pi$ appropriately
induces a Galois connection~\cite{cousot_galois}
between this abstract domain and the concrete heap domain.

For abstract commutativity to soundly entail concrete commutativity,
we establish a soundness relation
between concrete and abstract programs,
in the style of abstract interpretation~\cite{cousot-cousot-abstract-interp}.
It demands, for instance, that
$\ap{push}$ overapproximate $\tt{push}$
and that $\ap{push}$ fail when $\tt{push}$ fails on an in-purview heap.

{\bf Summary.} From our user-provided Stack abstraction, and typical abstract interpretation techniques, 
we can immediately derive sound heap commutativity conditions
from simple mathematical conditions:
\begin{center}
\begin{tabular}{rl|rl|l|l}
  \multicolumn{2}{c|}{Concrete} & \multicolumn{2}{c|}{Abstract} & Abstract & Derived \\
  \multicolumn{2}{c|}{Programs} & \multicolumn{2}{c|}{Functions} &  Condition & Concrete Condition \\
  \midrule
  %
  $\tt{push}$ & $\tt{push}$ & $\ap{push}(u)$ & $\ap{push}(v)$ & $ u=v$ &
    $\exists u.\, \PsiStack^+(\_) \ast a \mapsto u \ast b \mapsto u$ \\
  $\tt{push}$ & $\tt{pop}$ & $\ap{push}(u)$ & $\ap{pop}()$ & $ s=u::\_$ &
    $\exists u.\, \PsiStack^+(u\cons \_) \ast a \mapsto u \ast b \mapsto \_$ \\
  $\tt{pop}$ & $\tt{pop}$ 
  & $\ap{pop}()$ & $\ap{pop}()$ & $ s=x::x::\_$ &
    $\exists x.\, \PsiStack^+(x \cons x \cons \_) \ast a \mapsto \_ \ast b \mapsto \_$
\end{tabular}
\end{center}

In the remainder of this paper we formalize the above discussion,
share properties of commutativity,
and define composition on abstract domains.
Then in \S\ref{sec:full-examples}
we elaborate on the Stack example (adding a \tt{peek} operation)
along with other examples.
All definitions, lemmas, theorems, and examples have been mechanized 
in Coq (available in the \href{\supplementlink}{supplement}), except for those concerning the example 
concrete semantics in \S\ref{subsec:particular-semantics}.

%% file: conc_progs/stk-push-wide.tex
\begin{minipage}[t]{1.5in}%
  {\small\vspace{-\baselineskip-\abovedisplayskip}
  \begin{alignat*}{2}
    & \tt{push}_\ell^a \coloneqq 
    \langLet\ p = \langDeref \ell\ \langIn \\[-2pt]
    & \langInd \langLet\ v = \langDeref a\ \langIn \\[-3pt]
    & \langInd \langLet\ x = (v,p)\ \langIn \\[-3pt]
    & \langInd \langLet\ q = \langRef\ x \\[-3pt]
    & \langInd \langIn\ \ell \langWrite q
  \end{alignat*}}
\end{minipage}

%% file: conc_progs/stk-pop-wide.tex
\begin{minipage}[t]{1.9in}%
  {\small\vspace{-\baselineskip-\abovedisplayskip}
  \begin{alignat*}{2}
  & \tt{pop}_\ell^a \coloneqq 
  \langLet\ p = \langDeref \ell\ \langIn \\[-2pt]
  & \langInd \langIf\ p = \vnull\ \langThen\ \langFail\ \langElse \\[-3pt]
  & \langInd \langLet\ x = \langDeref p\ \langIn \\[-3pt]
  & \langInd \langLet\ v = \langFst\ x\ \langIn\ 
  \langLet\ q = \langSnd\ x \\[-3pt]
  & \langInd \langIn\ \ell \langWrite q\ \langSeq\
    \langFree\ p\ \langSeq\ 
  a \langWrite v
  \end{alignat*}}
\end{minipage}

%% file: figures/stk-abstr-overview.tex
$\begin{array}{ll}
  \pi(h) & \coloneqq s \ \text{if} \ \PsiStack^+(s,h) \ \text{for}\
  s \in \mathrm{list}(\mathrm{value})\ \text{else}\ \opurv, \ \text{where} \\[6pt]
  \PsiStack(s) & \coloneqq
    \exists p \in \mathrm{address}.\
    \ell \mapsto p \ast
    \psi(s,p) \\
  \psi(v \cons s, p) & \coloneqq
    \exists q \in \mathrm{address}.\
    p \neq \vnull \ast
    p \mapsto (v,q) \ast
    \psi(s,q) \\
  \psi([],p) & \coloneqq
    p = \vnull
\end{array}$

%% file: sections/prelim.tex

\section{Preliminaries}

Here we outline conventions and notations for heaps and heap predicates,
as well as our choice of concrete heap semantics.

\subsection{Heaps and Heap Predicates}
A \emph{heap} is a finite partial map from addresses (\L) to concrete values (\V).
Denote the set of all heaps with \H,
the set of non-null values with $\V^* = \V \setminus \set{\vnull}$,
and the finite address domain of a heap $h$ with $\domain{h}$.

Concrete values may be any typical datatypes
(integers, booleans, pairs, etc.) which are generally fixed in size
(e.g.\ an unbounded list would not typically be stored as one concrete value).
Addresses themselves can be treated as concrete values.
Assume unlimited addresses are available,
namely $\forall h.\, \domain{h} \subsetneq \L$.

Heaps $h_1$ and $h_2$ are \emph{disjoint} ($h_1 \disjoint h_2$)
iff $\domain{h_1} \disjoint \domain{h_2}$.
The \emph{union} of disjoint heaps ($h_1 \union h_2$)
yields the combined partial map of $h_1$ and $h_2$. 
Dually, a heap may be partitioned into smaller disjoint heaps.
$h_1$ is a \emph{subheap} of $h_2$ ($h_1 \subheap h_2$)
iff $h_1 \union k = h_2$
for some $k \disjoint h_1$.
Heap equality is defined extensionally.

A \emph{heap predicate} characterizes the addresses and values of heaps.
Given heap predicate $\Psi$ on \H,
$\Psi(h)$ denotes that $h$ satisfies $\Psi$, and
$\H(\Psi)$ is the set of all $\Psi$-satisfying heaps.
Sometimes we parameterize a heap predicate on some domain $V$.
In this case, $\Psi(v,h)$ denotes that $h$ satisfies $\Psi$ with $v \in V$.

Aside from heap predicates containing pure logical expressions and quantifiers,
we will describe heap contents directly 
with two standard predicates from separation logic \cite{reynolds_sep_logic,Parkinson}.
The \emph{points-to} predicate $p \mapsto v$,
for $p \in \L$ and $v \in \V$,
holds for $h$ iff 
$\domain{h} = \set{p}$ and $h(p) = v$.
The \emph{separating conjunction} operator $\Psi \ast \Phi$,
for predicates $\Psi$ and $\Phi$,
holds for $h$ iff $h = h_1 + h_2$ for some $h_1$ and $h_2$
s.t.\ $h_1 \disjoint h_2$ and $\Psi(h_1)$ and $\Phi(h_2)$.
$\ast$ is commutative and associative.

We will also want to characterize
when a \emph{portion} of a heap satisfies a particular predicate.
For this,
we define heap predicate \emph{extension}:
\begin{definition}
  \label{def:extension}
  For predicate $\Psi$,
  $h$ satisfies \emph{extension} $\Psi^+$ 
  iff a subheap of $h$ satisfies $\Psi$.
  Namely,
  $\Psi^+(h) \equiv (\exists \Phi.\, \Psi \ast \Phi)(h)$
  and  $\Psi^+(v,h) \equiv (\exists \Phi.\, \Psi(v,\cdot) \ast \Phi)(h)$.
\end{definition}

\subsection{Concrete Semantics}
\label{sec:concrete-semantics}

We choose to represent concrete programs
as total functions over heaps,
so that our semantics are highly general,
and thus broadly applicable.
We admit semantics which are nondeterministic, 
but must be constructive and must terminate. 
Nondeterminism mainly arises
during address allocation,
though this work is agnostic
to the source of nondeterminism.

\begin{definition}
  \label{def:conc-prog}
  A \emph{(concrete) program} is a
  total function $f : \H \ra \pset\H$,
  subject to the following properties:
  \begin{enumerate}
    \item On any initial heap $h$,
      $f$ terminates or fails in finite time.
     If $f$ successfully terminates on $h$, then $f(h)$ yields the set of
      all possible final heaps in which $f$ may terminate.
      If $f$ fails on $h$, then $f(h) = \emptyset$.
    \item $f$ reads any input arguments from,
      and writes any outputs to, fixed locations on the heap.
      We explore this in \S\ref{sec:full-examples}.
    \item $f$ \emph{acts locally}. Namely, 
      if $f(h) \neq \emptyset$, 
      then for any $h' \disjoint h$:
      $f(h \union h') \neq \emptyset$ and 
      $\forall k \in f(h \union h').\ h' \subseteq k$.
  \end{enumerate}
\end{definition}

\emph{Local action} enforces that $f$ does not modify data outside its footprint, as a weak form
of the typical frame rule of separation logic~\cite{reynolds_sep_logic}.
For any particular semantics that satisfies framing by construction,
the local action constraint should immediately hold for all concrete programs.
When working with compound abstractions,
we will define a stronger framing property (see Def.~\ref{def:capture}).

For two programs to commute,
they must terminate without failure in the first place.
To that end, we classify 
the initial heaps upon which $f$ can execute.
\emph{Sufficient} heaps contain the necessary addresses and appropriate values
for $f$ to run.
In contrast, heaps in the \emph{footprint} of $f$ contain appropriate addresses,
but may not contain appropriate values; thus
a heap in $f$'s footprint on which $f$ fails
cannot be ``corrected'' with further allocations.
Precisely:
\begin{align*}
  \sufficient{f} & \triangleq \setbuild{h}{f(h) \neq \emptyset} \\
  \footprint{f} & \triangleq \sufficient{f} \cup
    \setbuild{h}{\forall h' \disjoint h.\, h \union h' \notin \sufficient{f}}
\end{align*}

Our \emph{concrete domain}, as the counterpart
to our eventual abstract domain,
will be $\sC \triangleq \pset\H$.
\emph{Concrete transformers} over this domain map over sets of heaps. 
We derive these transformers directly
from programs:
\begin{definition}
  \label{def:conc-transformer}
  The \emph{concrete transformer} constructed from $f$, denoted $\bar f$, 
  is defined as $\bar f(C) = \bigcup_{h\in C} f(h)$ if $C \subseteq \sufficient{f}$,
  else $\emptyset$.
\end{definition}

\noindent
$\bar f$ yields the set of all possible output heaps given a set of inputs,
but fails entirely if $f$ fails on any individual input.
Relatedly, we define the composition/sequence of two programs so
that the second program fails entirely if it
fails on any one output of the first program,
namely $f;g \triangleq \bar g \circ f$.

\subsection{An Example Concrete Language}
\label{subsec:particular-semantics}

{

We share here a simple language called \lang, which
features allocation (\langRef), deallocation (\langFree), writing ($\langWrite$), reading (\langDeref), and branching.
We will use \lang\ in our example programs throughout the paper,
as an instance of the general semantics of \S\ref{sec:concrete-semantics}.
(N.B.\ \lang\ has not been formalized in Coq. However, 
all of our results regarding the general semantics
have been proved in Coq.)
\begin{alignat*}{2}
\stmt \Coloneqq&\ 
  \id \leftarrow \val \ | \
  \langFree\ \id \ | \
  \stmt \mathop{\langSeq} \stmt \ |\ \langFail\ | \ \langSkip \\
  | \ &\
  \langLet\ \id \coloneqq \expr\ \langIn\ \stmt 
  \ | \ 
  \langIf\ \val\ \langThen\ \stmt\ \langElse\ \stmt \\
\expr \Coloneqq&\ \langRef\ \val \ | \ \langDeref (\id \ |\ \L) 
  \ | \
  \val + \val \ | \ \val = \val \ | \ \cdots \\
\val \Coloneqq&\ \id \ |\ \V
\end{alignat*}

\noindent
\lang\ has a nondeterministic allocation procedure:
for any initial heap $h$,
\langRef\ yields an address from a set $\sf{fresh}(h)$
s.t.\ $\sf{fresh}(h) \subseteq \L \setminus \domain{h}$.
All other terms are deterministic.
We elide the remaining details of \lang's semantics,
which are typical for an imperative heap language.
When working with \lang\ programs throughout the paper,
we implicitly lift them to the concrete programs 
of Def.~\ref{def:conc-prog}.

%% file: sections/define.tex

\section{Abstract Domain}
\label{sec:construction}

In this section we formalize our abstract domain
and abstract programs operating within this domain,
and establish a soundness relationship between concrete and abstract programs.
This will let us perform commutativity reasoning easily with abstract objects,
and obtain concrete commutativity results for free.

\subsection{Constructing the Domain}
\label{subsec:contruct-abstraction}

We design our abstract domain to lift
heap-allocated structures to mathematical objects.
Each instance of an abstraction consists of 
a set of \emph{abstract values} (e.g. integers, sets, sequences)
and a mapping from heaps into said values.
A heap with an allocated structure appropriate
to be mapped to an abstract value
is said to be \emph{``within the purview''}
of a given abstraction.
On the other hand, poorly structured heaps
are \emph{``outside the purview''} of the abstraction,
and get mapped to a special abstract value, $\opurv$ (``cross'').

When constructing an abstraction,
we impose a \emph{locality constraint}
which demands that information extracted from heaps be finite in scope,
so that any heap \emph{containing} a valid structure
is in-purview.
Some results of this work actually do not depend on this constraint.
However, it is a valuable property to strive for,
and indeed it should hold for most typical data structure abstractions.

The abstract domain is defined formally as follows:
\newcommand*{\po}{\leq_\aA}

\begin{definition}
  \label{def:abstraction}
  Build an abstraction $\aA = \abstr{X}{\pi}$
  from a set of \emph{abstract values}
  $X = \set{x_1,x_2,\dots}$,
  and a \emph{projection function} $\pi:\H\ra X+\opurv$
  subject to the \emph{locality constraint} that
  for any disjoint $h$ and $h'$,
  $\pi(h) \in X \Rightarrow \pi(h \union h') = \pi(h)$.
  The full domain
  is $\sA_\aA \triangleq X{+}\bot{+}\top{+}\opurv{+}\ipurv$,
  with partial ordering $\leq_\aA$,
  where
  $\bot \po \opurv \po \top$
  and
  $\forall x \in X.\ \bot \po x \po \ipurv \po \top$
  and $\forall x_1,x_2 \in X.\ x_1 \po x_2 \Leftrightarrow x_1 = x_2$.
\end{definition}

\noindent
Denote all in-purview heaps of \aA\
with $\purview\aA \triangleq \setbuild{h}{\pi(h) \in X}$,
and denote its abstract values and projection function
with $X_\aA$ and $\pi_\aA$.

$\sA_\aA$ is a complete lattice,
and is thus equipped with
a typical
join ($\sqcup$) operation.
Recalling from \S\ref{sec:concrete-semantics}
that our concrete domain is $\sC = \pset\H$,
we connect the concrete and abstract domains as follows:
\begin{definition}
\label{def:alpha-gamma}
Define \emph{abstraction} and \emph{concretization} functions
$\alpha_{\sf A} : \sC \ra \sA_\aA$ and
$\gamma_{\sf A} : \sA_\aA \ra \sC$,
where 
$\alpha_{\sf A}(C) =
\bigsqcup_{h \in C} \pi(h)$
and
$
\gamma_{\sf A}(x) =
\setbuild{h \in \H}{\pi(h) \leq_{\sf A} x}$.
\end{definition}

\noindent
$\alpha_{\sf A}$ and $\gamma_{\sf A}$ are each monotone,
and together they form a Galois connection~\cite{cousot_galois} which
induces meanings for values in $\sA_\aA$:
$\bot$ is the failure state,
$x_i \in X$ abstracts sets of in-purview heaps which all map to $x_i$,
$\ipurv$ abstracts any set of in-purview heaps,
$\opurv$ out-of-purview heaps,
and $\top$ any heaps.

We now share three basic examples of abstract domains.
In subsequent sections 
we will 
reason about concrete and abstract programs which operate
on these defined structures.
In \S\ref{sec:full-examples} we explore each example in full.

\begin{example}[Non-negative Counter]
\label{ex:ctr}

A ``hello world'' data structure in commutativity
is the non-negative counter (NNC)---an integer which may be
incremented, decremented (but not below 0), and read.
We abstract an NNC located at address $p$ as 
$\ctr_p \coloneqq \abstr{\N}{\pi}$, where
$\pi(h) \coloneqq n$ if $(p \mapsto n)^+(h)$ for $n \in \N$,
else $\opurv$.
The \emph{extension} (Def.~\ref{def:extension}) on the predicate $(p \mapsto n)$
ensures that $\ctr_p$ satisfies locality,
allowing the domain to capture
any heap \emph{containing} a counter. 

\end{example}

\begin{example}[Two-Set]
\label{ex:two-set}

Consider an unordered set
containing at most two elements of type $T \subseteq \V^*$.
These elements are stored at fixed addresses $p$ and $q$,
with $\vnull$ otherwise stored as a placeholder.
One might construct $X$ and $\pi$ for this structure in a few equivalent ways,
the key property being that concrete two-sets containing
the same elements should
map to the same abstract value,
regardless of their order.
In the following approach,
the abstract values are simply mathematical sets
of size at most two.
We define $\twoset_{p,q}^T \coloneqq \abstr{
  \setbuild{S \in \pset T}
    {\abs{S} \leq 2} 
}{\pi}$ with:
\input{figures/ts-abstr.tex}

Note how we use $\PsiSet^+$ throughout $\pi$
to satisfy locality.
One technicality is ensuring
that $\pi$ is well-defined,
namely that $\pi(h)$ yields one unambiguous abstract value for each $h$.
In Coq we prove that this $\pi$ is indeed well-defined.

\end{example}

\begin{example}[Linked Stack]
\label{ex:stack-abstraction}

Reiterating our overview example (\S\ref{sec:overview}),
consider a stack 
implemented on the heap as a linked list
and accessed at fixed address $\ell$.
Constructing new cells when pushing to the stack involves nondeterministic address allocation;
two stacks containing the same values but
different intermediate addresses should be considered observationally equivalent.
To achieve this, our recursive construction of $\pi$
will existentially quantify intermediate addresses,
and not lift them explicitly into the abstract domain.
We define
$\sf{Stk}_\ell \coloneqq \abstr{\mathrm{list}(\V^*)}{\pi}$ with
\begin{center}
  \input{figures/stk-abstr.tex}
\end{center}

\noindent
Again, we must show that each $h$ satisfies $\PsiStack^+(s,h)$ for at most one $s \in \mathrm{list}(\V^*)$. %
\end{example}

\subsection{Isomorphism of Abstract Domains}
\label{subsec:iso}

A notion of isomorphism between abstract domains
will also prove useful:
\begin{definition}
  \label{def:induce-iso}
  A bijection $\varphi : X_\sf A \ra X_\sf B$
  induces an isomorphism between \sf A and \sf B,
  denoted $\sf A \cong_\varphi \sf B$, if
  $\forall h.\ \varphi(\pi_\sf A (h)) = \pi_\sf B (h)$.
  \sf A and \sf B are \emph{isomorphic} $(\sf A \cong \sf B)$
  if such a bijection exists.
\end{definition}
This definition is 
stronger than necessary for some of our results;
future work may warrant distinguishing isomorphism
from a weaker equivalence 
which relaxes the bijectivity requirement of $\varphi$.
See the \ARXIV{\hyperref[appdx:isomorphism]{Appendix}}%
\VMCAI{appendix of~\cite{technicalreport}}%
for an example
where we construct a two-set abstraction which is isomorphic
to that of Ex.~\ref{ex:two-set}.

%% file: figures/ts-abstr.tex
\begin{gather*}
  \PsiSet\big((u,v),\cdot\big) \coloneqq
    p \mapsto u \ast
    q \mapsto v \hspace*{2em} \\
  \pi(h) \coloneqq
    \begin{cases}
      \set{u,v} &
        \PsiSet^+\big((u,v),h\big) \wedge
        u \neq v\ \text{for}\
        u,v \in T \\
      \set{u} &
        \PsiSet^+\big((u,\vnull),h\big) \vee
        \PsiSet^+\big((\vnull,u),h\big)\
        \text{for}\ u \in T \\
      \emptyset &
        \PsiSet^+\big((\vnull,\vnull),h\big) \\
      \opurv & \text{else}
    \end{cases}
\end{gather*}

%% file: figures/stk-abstr.tex
$\begin{array}{ll}
  \pi(h) & \coloneqq s \ \text{if} \ \PsiStack^+(s,h) \ \text{for}\
  s \in \mathrm{list}(\V^*)\ \text{else}\ \opurv, \ \text{where} \\[6pt]
  \PsiStack(s) & \coloneqq    
    \exists p \in \L + \vnull.\
    \ell \mapsto p \ast
    \psi(s,p) \\
  \psi(v \cons s, p) & \coloneqq
    \exists q \in \L + \vnull.\
    p \neq \vnull \ast
    p \mapsto (v,q) \ast
    \psi(s,q) \\
  \psi([],p) & \coloneqq
    p = \vnull
\end{array}$

%% file: sections/semantics.tex

\subsection{Abstract Semantics and Soundness}

We design our abstract semantics so that
concrete heap behavior (valid or invalid) can be threaded up to the abstract domain 
with enough precision that,
later, abstract commutativity will soundly entail concrete commutativity~(\S\ref{sec:commutativity}).

An \emph{abstract program}
in the context of abstraction $\aA = \abstr X \pi$
is any function from $X$ to $\sA$,
from which we derive an \emph{abstract transformer} over \sA\
(recall from Def.~\ref{def:abstraction} that
$\sA = X + \bot + \top + \opurv + \ipurv$).
While these transformers could be constructed
in various ways, for our purposes
the following will suffice:
\begin{definition}
  $\widehat m$ denotes the abstract transformer derived
  from $m : X \ra \sA$, where
  $\widehat m(x) = \bot$ for $x = \bot$,
  $m(x)$ for $x \in X$,
  and $\top$ otherwise.
\end{definition}

To relate the behavior of an abstract program $m$ defined in \aA\
to a concrete program $f$,
we establish a notion of \emph{soundness}
(recall $\alpha$ from Def.~\ref{def:alpha-gamma}):
\begin{definition}
  \label{def:soundness}
  $m$ \emph{soundly abstracts} $f$ \emph{in} $\aA$,
  denoted $\sound{\aA}{m}{f}$,
  if
  \begin{enumerate}
    \item $
      \forall C, x. \enspace
      \alpha_\aA (C) \leq_\aA x \; \implies \;
      \alpha_\aA \left(\bar f (C)\right) \leq_\aA \widehat m (x)
    $,\enspace and
    \item $
      \forall C. \enspace
      \alpha_\aA (C) \in X \; \wedge \;
      \alpha_\aA \left(\bar f (C)\right) = \bot \; \implies \;
      \widehat m (\alpha_\aA (C)) = \bot.
    $
  \end{enumerate}
\end{definition}

\noindent
The first condition is typical for abstract interpretation,
demanding that $m$ \emph{overapproximate} $f$.
This lets us be
imprecise with the definition of abstract programs if desirable
(e.g.\ to avoid complex behaviors),
the trade-off being that concrete commutativity conditions
eventually derived may yield false negatives 
(i.e.\ not be the weakest precondition).
The second soundness condition demands that $m$ fail (yield $\bot$)
whenever $f$ fails on an in-purview input.
Because the failure sink state is at the bottom of the abstract lattice
(Def.~\ref{def:abstraction}),
without this condition there could be scenarios
where two abstract programs successfully commute,
but their corresponding concrete programs fail to execute.

We now return to the example of the non-negative counter
and its abstraction $\ctr_p$ (see Ex.~\ref{ex:ctr}),
to define the corresponding concrete and abstract programs
for increment and decrement.
We cannot define \emph{read} in $\sf{Ctr}_p$,
because \emph{read} must write its output to an additional location on the heap.
We similarly must wait to revisit 
our two-set and linked stack examples,
whose methods all have inputs and/or outputs.
We develop the machinery needed for inputs and outputs
in \S\ref{sec:conjunction},
and reason fully about all three examples in \S\ref{sec:full-examples}.

\begin{example}[Non-negative Counter]
\label{ex:semantics-nnc}

Consider an NNC allocated at address $p$.
We begin with concrete implementations of increment and decrement,
written in the \lang\ language (\S\ref{subsec:particular-semantics}),
and implicitly lift them to concrete programs.
\begin{center}
  \input{conc_progs/ctr-incr-wide.tex}
  \input{conc_progs/ctr-decr-wide.tex}
\end{center}

\noindent
Next, defining abstract increment and decrement functions in $\sf{Ctr}_p$
is intuitive:
\begin{equation*}
  \ap{incr}_p(n) \coloneqq n + 1 \hspace*{4em}
  \ap{decr}_p(n) \coloneqq
    \max(0,n-1)
\end{equation*}
We can show  that
$\sound{\ctr_p}{\ap{incr}_p}{\hspace*{-0.1em}\tt{incr}_p}$
and $\sound{\ctr_p}{\ap{decr}_p}{\hspace*{-0.1em}\tt{decr}_p}$
with typical symbolic execution techniques.
Had we defined \tt{decr} to fail when the counter is~0,
rather than skip,
we would have to define $\ap{decr}_p(0) = \bot$
to maintain soundness.

\end{example}

%% file: conc_progs/ctr-incr-wide.tex
\begin{minipage}[t]{1.8in}%
  {\small\vspace{-\baselineskip-\abovedisplayskip}
  \begin{alignat*}{2}
    & \tt{incr}_p \coloneqq 
    \langLet\ c = \langDeref p\ \langIn \\[-2pt]
    & \langInd \langLet\ i = c + 1\ \langIn\ 
    p \langWrite i
  \end{alignat*}}
\end{minipage}

%% file: conc_progs/ctr-decr-wide.tex
\begin{minipage}[t]{2.5in}%
  {\small\vspace{-\baselineskip-\abovedisplayskip}
  \begin{alignat*}{2}
    & \tt{decr}_p \coloneqq 
    \langLet\ c = \langDeref p\ \langIn\ 
    \langLet\ i = c - 1\ \langIn \\[-3pt]
    & \langInd \langIf\ i < 0\ \langThen\ \langSkip\ 
    \langElse\ p \langWrite i
  \end{alignat*}}
\end{minipage}

%% file: sections/commutativity.tex
\section{Sound Commutativity}
\label{sec:commutativity}

With our concrete and abstract semantics established,
we now define commutativity in the concrete and abstract spaces.
We then relate the two with a soundness theorem that
reduces concrete commutativity reasoning to
simpler reasoning
about abstract programs.

\subsection{Defining Commutativity}

We say concrete programs $f$ and $g$ \emph{commute},
under a notion of observational equivalence on a subset of all heaps,
when $f;g$ and $g;f$ both terminate successfully
and yield observationally equivalent outcomes.
We denote an obs.\ eq.\ relation with $[\Psi,\sim]$,
for a predicate $\Psi$ on \H\
and an eq.\ relation $\sim$ on $\H(\Psi^+)$.
Then concrete commutativity is defined precisely as:

\newcommand*{\commuteobseq}[5]{#3 \bowtie^{#4}_{[#1,#2]} #5}

\begin{definition}
  For \emph{commutativity condition} $P$ on \H,
  $f$ and $g$
  \emph{commute under $P$ w.r.t.\ ${[\Psi,\sim]}$},
  denoted $\commuteobseq\Psi\sim f P {\hspace*{-0.1em} g}$, if
  $\forall h \in \H(P)$.\
  $(f;g)(h), (g;f)(h) \neq \emptyset$ and
  $(f;g)(h) \cup (g;f)(h) \subseteq [h']_\sim$ 
  for some $h' \in \H(\Psi^+)$.
\end{definition}

\noindent
The eq.\ relation we use for concrete commutativity is often very similar
to the relation implicitly induced by an abstraction.
We can make this explicit by deriving a concrete relation in terms of an abstraction,
a technique we will use extensively in the remainder of the paper.

\begin{definition}
  \label{def:obseq-from-abstr}
  For abstraction $\sf A$,
  define concrete equivalence $[\purview{\sf A},\sim_\aA]$ with
  $h \sim_\aA h' \coloneqq \pi_\aA(h) = \pi_\aA(h')$.
  Use the shorthand $\commuteconc{\aA}{f}{P}{\hspace*{-0.1em} g}$
  for $\commuteobseq{\purview{\aA}}{\sim_\aA}{f}{P}{\hspace*{-0.1em} g}$.
\end{definition}

Naturally, a user is responsible for choosing an observational equivalence relation
(or in light of Def.~\ref{def:obseq-from-abstr}, an abstraction) which is sufficiently descriptive for their purposes.
Properties of concrete commutativity include:
\begin{mathpar}
\inferrule{
  {\begin{gathered}
    \H(P') \subseteq \H(P) \\[-4pt]
    \commuteconc{[\Psi,\sim]}{f}{P}{\hspace*{-0.1em} g}
  \end{gathered}}
}{
  \commuteconc{[\Psi,\sim]}{f}{P'}{\hspace*{-0.2em} g}
}
\and
\inferrule{
  {\begin{gathered}
    \sf A \cong \sf B \\[-4pt]
  \commuteconc{\sf A}{g}{P}{f} 
  \end{gathered}}
}{
  \commuteconc{\sf B}{g}{P}{f}
}
\and
\inferrule{
  \commuteconc{[\Psi,\sim]}{f}{P}{g}
}{
  \commuteconc{[\Psi,\sim]}{g}{P}{f}
}
\and
\inferrule{
  {\begin{gathered}
    \forall h,h'.\ h \sim h' \implies h \sim' h' \\[-4pt]
  \commuteconc{[\Psi,\sim]} f P g 
  \end{gathered}}
}{
  \commuteconc{[\Psi',\sim']} f P g
}
\end{mathpar}

Next we define \emph{abstract commutativity},
in a manner 
so that we can achieve a useful soundness guarantee
in relation to concrete commutativity.
\begin{definition}
  \label{def:commute-abstract}
  Abstract programs $m$ and $n$ defined in \aA\
  \emph{commute under} predicate $Q$ on $X_\aA$, 
  denoted $\commuteabs\aA m Q n$, if
  $\forall x \in X_\aA.\,
  Q(x) \Rightarrow
  \widehat n(m(x)) = 
  \widehat m(n(x)) \in X_\aA$.
\end{definition}
\noindent
Note how we only accept outcomes within the purview of \aA.
We reject $\bot$, as non-failure
is a prerequisite for commutativity.
We also reject $\opurv$, $\ipurv$, and $\top$, as
such outcomes are not precise enough
to guarantee meaningful commutativity.

\subsection{Sound Commutativity Theorem}

To relate abstract and concrete commutativity,
we must first relate the relevant abstraction and observational equivalence relation:

\begin{definition}
$\aA$ \emph{captures} $[\Psi,\sim]$ iff
$\H(\Psi^+)\subseteq\purview\aA$ and
\\
\DefineIndent
$\forall h,h' \in \H(\Psi).\, \pi_\aA(h) = \pi_\aA(h') \in X_\aA \implies h \sim h'$.
\end{definition}

\noindent
This notion of \emph{capture} is preserved by abstraction isomorphism.
Additionally, $\aA$ always captures $[\purview{\sf A},\sim_\aA]$,
a fact we will frequently use implicitly.

We now present the central result of our work,
the \emph{sound commutativity theorem},
which offloads the verification (via symbolic execution)
of concrete commutativity conditions
to much simpler reasoning about abstract programs:

\begin{theorem}
  \label{thm:commute-sound}
  If $\sound{\sf A}{m}{f}$, $\sound{\sf A}{n}{g}$,
  $\commuteabs{\sf A}{m}{Q}{\hspace*{-0.1em} n}$,
  and $\aA$ captures $[\Psi,\sim]$,
  then \\
  \DefineIndent 
  $\commuteconc {[\Psi,\sim]} f {P^+}  {\hspace*{0em}g}$,
  where $P(h) \equiv \pi_\aA(h) \in X_\aA \wedge Q(\pi_\aA(h))$.
\end{theorem}

\noindent
Sound commutativity takes a valid abstract commutativity condition
and transforms it automatically into
a concrete condition.
For instance, suppose $\pi_\aA$ takes the fairly common form,
for some $\Phi$,
of
$\pi_\aA(h) = x \ \text{if}\ \Phi(x,h)\ \text{for}\ x \in X_\aA, \text{else}\ \opurv$.
In this case, the concrete condition
we derive from abstract condition $Q$ is
$\exists x \in X.\ \Phi^+(x,h) \ast Q(x)$.
Indeed, any condition we derive with 
Thm.~\ref{thm:commute-sound}
is \emph{extended} (Def.~\ref{def:extension}).
This provides \emph{framing} for free---%
it guarantees if $f$ and $g$ commute in some heap context,
then they also commute in any larger context.

With sound commutativity established,
we can summarize the overall pattern of reasoning:
(i) construct an abstract domain,
(ii) define abstract programs for each concrete program
  and prove soundness between them using symbolic execution,
(iii) verify or synthesize commutativity conditions for abstract program pairs,
and (iv) immediately derive concrete conditions.

\begin{example}[Non-negative Counter]
  \label{ex:nnc-commute}

Returning to the NNC programs
defined in Ex.~\ref{ex:semantics-nnc},
we can perform abstract commutativity reasoning easily.
For instance, 
to verify that $\commuteabs{\sf{Ctr}_p}{\ap{incr}_p}{Q}{\hspace*{-0.2em} \ap{decr}_p}$
for $Q(n) \equiv n > 0$,
we simply evaluate:
\newcommand*{\fincr}{\widehat{\ap{incr}}_p}
\newcommand*{\fdecr}{\widehat{\ap{decr}}_p}
\begin{displaymath}
  \fdecr(\ap{incr}_p(n)) = \fdecr(n+1) = n
  \hspace*{2em}
  \fincr(\ap{decr}_p(n)) = \fincr(n-1) = n
\end{displaymath}

\noindent
With proofs of soundness (Def.~\ref{def:soundness}) 
of $\ap{incr}$ 
and $\ap{decr}$ w.r.t. \tt{incr} and \tt{decr},
we immediately derive via Thm.~\ref{thm:commute-sound}
that $\commuteconc {\sf{Ctr}_p} f {P^+} {\hspace*{-0.1em}g}$
where $P \equiv \exists n \in \N^+.\ p \mapsto n$.
Note how we use the equivalence imposed by $\ctr_p$ as
our concrete observational eq.\ relation (Def.~\ref{def:obseq-from-abstr}).
We might \emph{synthesize}
an abstract condition interactively
(we leave automation to future work),
by trying to prove that
$\ap{incr}$ and $\ap{decr}$ commute under $\top$,
and seeing where the proof gets stuck.
In this case,
we will get stuck showing that $1 = 0$
under an initial abstract value of $0$.
By constraining that $n > 0$,
the proof can be completed;
thus $n > 0$ is a valid precondition.

\end{example}

%% file: sections/conjunction.tex
\section{Abstract Domain Composition}
\label{sec:conjunction}

In this section, we expand upon our established tools for abstracting data structures and performing commutativity analysis,
by defining composition of abstract domains.
This enables compositional reasoning in settings with multiple data structures,
and provides machinery needed to reason about
concrete programs with heap-allocated inputs and outputs.
Various compositional operators over abstract domains have been used in abstract interpretation,
such as Cartesian product, reduced product, and reduced power \cite{abstr-interp-products-survey,Giacobazzi1,Giacobazzi2,Giacobazzi3}.
While such an existing operator is likely suitable,
we elect to design our own simple operator.

A composition of abstractions should yield a new abstraction,
and should respect
the disjointness of the concrete structures
which are individually abstracted.
We achieve this with the
\emph{abstract conjunction} operator,
denoted $\sf A \ast \sf B$,
which is defined to
mirror the concrete separating conjunction.
The purview of $\sf A \ast \sf B$ will contain heaps
which can be partitioned into disjoint halves that
are respectively in the purviews of \sf A and \sf B.
\begin{definition}
  \label{def:conjunction}
  $\sf A \ast \sf B \triangleq \abstr{X_{\sf A} \times X_{\sf B}}{\pi}$, where
  $\pi(h) = (\pi_{\sf A}(h_1),\pi_{\sf B}(h_2))$\\
  \DefineIndent
  if $h = h_1 + h_2$ for some $h_1 \disjoint h_2$ s.t.\
  $h_1 \in \purview{\sf A}$ and $h_2 \in \purview{\sf B}$,
  else $\opurv$.
\end{definition}

A technicality of Def.~\ref{def:conjunction} is ensuring that
$\pi(h)$ maps to a \emph{unique} $(a,b)$ pair (or to \opurv) for every $h$.
We show in Coq that this uniqueness does hold,
due to
the locality constraint of Def.~\ref{def:abstraction}.
Additionally,
when \aA\ and \sf B satisfy locality,
so too does $\aA \ast \sf B$.
Abstract conjunction is compatible with,
and is commutative and associative up to, abstraction isomorphism;
we use these facts implicitly throughout \S\ref{sec:full-examples}.
Much like the heap separating conjunction,
$\sf A \ast \sf B$ may be \emph{unsatisfiable},
i.e.\ $\pi_{\sf A \ast \sf B}$ maps all heaps to \opurv.
Commutativity reasoning w.r.t.\ an unsatisfiable domain is still valid,
albeit meaningless,
as there are no in-purview heaps.
Applying the concrete observational equivalence construction
of Def.~\ref{def:obseq-from-abstr} to composition,
we get $[\purview{\aA \ast \aB}, \sim_{\aA \ast \aB}]$
where $h \sim_{\aA \ast \aB} h'$
if $h = h_a + h_b$ and $h' = h_a' + h_b'$ s.t.\
$h_a \sim_\aA h_a'$ and $h_b \sim_\aB h_b'$.

\subsection{Concrete Program Capture}
\label{sec:capture}

Effective compositional reasoning
will often require that behavior of
the concrete programs of interest
are ``captured'' by the abstract domains in use.
For instance, \tt{incr}
has well-defined behavior w.r.t.\ $\ctr_p$,
but not 
$\sf{Stk}_\ell$.
If we prove properties about \tt{incr}
w.r.t.\ $\ctr_p$,
we should be able to derive how \tt{incr} behaves w.r.t.\ $\ctr_p \ast \sf{Stk}_\ell$ (namely it leaves the stack untouched).
However, if we reason about \tt{incr} w.r.t.\ $\sf{Stk}_\ell$,
we will learn nothing about
how \tt{incr} behaves
w.r.t.\ $\sf{Stk}_\ell \ast \ctr_p$.

To this end,
we say that an abstraction \sf A \emph{captures} program $f$ if
(1) heaps in the purview of \sf A are in the footprint of $f$,
(2) $f$ maps in-purview heaps to in-purview heaps,
and
(3) $f$ maintains its specific in-purview mappings 
when any disjoint heaps are joined onto the initial heap
(thus strengthening \emph{local action} in an abstraction-specific manner).
Precisely:
\begin{definition}
  \label{def:capture}
  \aA\ \emph{captures} concrete program $f$
  if: 
  \begin{enumerate}
    \item 
      $\purview\aA \subseteq \footprint{f}$, and
    \item 
      $\forall h \in \purview\aA \cap \sufficient{f}.\
      \alpha_\aA (f(h)) \in X_\aA$, and
    \item 
      $\forall h \in
        \purview\aA \cap \footprint{f}, h' \disjoint h.$ 
      $\alpha_\aA \big(\mathopen{}\setbuild{k \heapdiff h'}{k \in f(h \union h')}\mathclose{}\big) =
       \alpha_\aA (f(h))$,\\
      where $k-h'$ is the subheap of $k$
      with domain $\domain k \setminus \domain {h'}$.
  \end{enumerate}
\end{definition}
\noindent
Proving capture can be less burdensome
than it appears---if we have shown that 
$\sound{\sf A}{m}{f}$
for $m$ s.t.\ $\mathsf{image}(m) \subseteq X_\aA + \bot$,
then conditions (1) and (2) immediately hold for $f$.
We also expect that (3) should hold automatically
for all abstractions and all \lang\ programs
(and other typical heap languages).
Capture is preserved by abstraction isomorphism;
and, if \aA\ captures $f$
then $\aA \ast \sf B$ captures $f$.

Capture yields
a common form of concrete commutativity:
\emph{commutativity from noninterference}.
If two programs
operate on disjoint structures,
then they automatically commute whenever they individually
terminate without failure:

\begin{theorem}
  \label{thm:commute-noninterfere}
  If \aA\ and \sf B capture $f$ and $g$ respectively,
  then \\
  \hspace*{2em}
  ${\commuteconc{\sf A \ast \sf B}{f}{P^+}{g}}$
  where $P(h) \equiv h \in \sufficient{f} \cap \sufficient{g}
          \cap \purview{\sf A \ast \sf B}$.
\end{theorem}

Finally, we will often discuss program capture and soundness in the same breath,
so we provide a convenient shorthand:
%
  $\soundcapt{\sf A}{m}{f}$ denotes that
  $\sound{\sf A}{m}{f}$
  and $\sf A$ captures $f$.

\subsection{Compositional Commutativity}

In service of sound commutativity reasoning in compound abstract domains,
we define a way to compose abstract programs:

\begin{definition}
    Given $m$ and $n$ defined in \sf{A} and \sf{B} respectively,
    their \emph{conjunction} is $m \ast n$ in $\sf{A} \ast \sf{B}$,
    where for $a \in X_\aA$ and $b \in X_{\sf B}$,
    $(m \ast n)(a,b) = \bot$ if $m(a) = \bot \vee n(b) = \bot$,
    $(m(a),n(b))$ if $m(a) \in X_\aA \wedge n(b) \in X_{\sf B}$,
    else $m(a) \sqcup n(b)$.
\end{definition}

The \emph{compound program soundness theorem}
then states that the conjunction of two abstract programs
soundly abstracts the \emph{sequence} of two concrete programs (in either order),
so long as these concrete programs operate on disjoint structures
(via \emph{capture}).
We also derive an \emph{abstract frame rule} of sorts.
\begin{theorem}
  \label{thm:conjoin-soundness}
  If $\soundcapt{\sf A} m f$ and $\soundcapt{\sf B} n g$, 
  then $\soundcapt{\sf A \ast \sf B}{m \ast n}{f;g}$
  and $\soundcapt{\sf A \ast \sf B}{m \ast n}{g;f}$.
\end{theorem}
\begin{corollary}
  \label{thm:abstract-frame-rule}
  If $\soundcapt \aA m f$,
  then
  $\soundcapt{\aA\ast\sf{B}}{m \ast \mathrm{id}}{f}$.
\end{corollary}

With composition of abstract domains and their programs
established,
we can demonstrate new compositional
commutativity soundness properties,
beginning with \emph{compound abstract commutativity}:

\begin{lemma}
  \label{lem:commute-conj}
  If $\commuteabs{\sf{A}}{m}{Q}{n}$
  and $\commuteabs{\sf{A'}}{m'}{Q'}{n'}$,
  then
  $\commuteabs{\sf{A}\ast\sf{A'}}{m\ast m'}{P}{n\ast n'}$ \\
  \DefineIndent
  where $P(a,b) \equiv Q(a) \wedge Q'(b)$.
\end{lemma}

\noindent
From this, along with
sound commutativity (Thm.~\ref{thm:commute-sound})
and compound program soundness (Thm.~\ref{thm:conjoin-soundness}),
we yield
\emph{compound sound commutativity},
which allows us to compose individual commutativity results
about pairs of programs operating on disjoint structures:

\begin{theorem}
  \label{cor:separable-commutativity-conjunction}
  If $\commuteabs{\sf A} m Q {m'}$, $\commuteabs{\sf B} n {Q'} {\hspace*{-0.1em}n'}$,
  $\soundcapt{\sf A} m f$,
  $\soundcapt{\sf A}{m'}{f'}$,
  $\soundcapt{\sf B} n g$,
  and $\soundcapt{\sf B}{n'}{g'}$,
  then $\commuteconc{\sf A \ast \sf B}{f;g}{P^+}{f';g'}$
  where
  $P(h) \equiv \pi_{\sf A \ast \sf B}(h) = (a,b) \in X_{\sf A \ast \sf B}
    \wedge Q(a) \wedge Q'(b)$.
\end{theorem}

\noindent
Note how this result 
still holds even if $\sf A \ast \sf B$ is unsatisfiable,
because the concrete commutativity condition requires
that the initial heap maps to a valid abstract value.
From this theorem 
we derive
\emph{framed sound commutativity}:

\begin{corollary}
  \label{cor:frame-commute}
  If $\soundcapt{\sf{A}}{m}{f}$
  and $\soundcapt{\sf{A}}{n}{g}$ 
  and $\commuteabs{\sf{A}}{m}{Q}{n}$,
  then $\commuteconc{\sf{A}\ast\sf{B}}{f}{P^+}{g}$,
  where $P(h) \equiv h \in \purview{\sf A \ast \sf B} \wedge
  Q(\mathrm{fst}(\pi_{\sf A \ast \sf B}(h)))$.
\end{corollary}

Let us contrast this result with our prior
sound commutativity result
(Thm.~\ref{thm:commute-sound}), which also
includes a form of framing.
Thm.~\ref{thm:commute-sound} says that,
\emph{within} the parameters of a
chosen abstraction and concrete observational eq.\ relation,
the derived concrete commutativity condition is framed.
On the other hand, Cor.~\ref{cor:frame-commute}
guarantees---%
under the stronger premise of \emph{capture}---%
that the programs commute under a framed concrete condition within \sf A,
as well as within the purview and observational equivalence
imposed by any compound abstraction \emph{containing} \sf A.

%% file: sections/full-examples.tex
\section{Examples in Full}
\label{sec:full-examples}

In this section we reiterate and expand upon our working examples:
a non-negative counter, 
a two-set, 
a linked-list stack, 
and the composition of a stack and counter.
For each example, we
(i) define concrete programs;
(ii) construct an abstract domain and abstract programs;
(iii) perform abstract commutativity reasoning;
(iv) prove soundness between the concrete and abstract programs;
and (v) derive concrete commutativity conditions.
As (v) follows directly from (iii) and (iv),
we sometimes omit it for brevity.

Throughout these examples,
we will need to abstract the
arguments and return values of methods.
To that end we define the ``address domain'' \sf{Addr},
which simply captures the value stored at a given heap address.
For address $p$ and values $V \subseteq \V$ considered valid,
construct
$\sf{Addr}_p^V \coloneqq \abstr{V}{\pi}$,
where $\pi(h) \coloneqq v$ if $(p \mapsto v)^+(h)$ for $v \in V$, else $\opurv$.

We will also use some shorthands. 
We may omit abstract program compositions
rendered trivial by (Cor.~\ref{thm:abstract-frame-rule}), 
e.g.\ if $m$ is defined in $\sf A$
and we are reasoning in $\sf A \ast \sf B$,
we may write $m$ rather than $m \ast \mathit{id}$.
Furthermore, we will leverage the associativity and commutativity
of abstract composition up to isomorphism 
to reason about abstract programs
which have been defined in a similar but 
``out-of-order'' compound abstraction.
%
Namely, we use the property that if
$\sound \aA m f$ and $\aA \cong_\varphi \sf B$,
then $\sound{\sf B}{\varphi \circ m \circ \varphi^{-1}} f$.
The addresses on which an abstract program is parameterized
will serve as unique identifiers for
how the input and output tuples of said program
should be implicitly permuted.

\subsection{Non-negative Counter}
\label{subsec:full-ex-nnc}

Consider a non-negative counter structure allocated at $p$,
featuring the following concrete operations written in \lang:
\begin{center}
  \input{conc_progs/ctr-incr.tex}
  \input{conc_progs/ctr-decr.tex}
  \hspace*{-0.8em}
  \input{conc_progs/ctr-read.tex}
\end{center}
\noindent
Note how \code{read} writes its output to a location $r$.
Define
$\ctr_p \coloneqq \abstr{\N}{\pi}$, where
$\pi(h) \coloneqq n$ if $(p \mapsto n)^+(h)$ for $n \in \N$,
else $\opurv$.
Next construct the abstract $\ap{incr}$ and $\ap{decr}$ programs in
$\ctr_p$,
and $\ap{read}$ in $\aA = \sf{Ctr}_p \ast \sf{Addr}_r^\N$
to accommodate its output:
\begin{gather*}
  \ap{incr}_p(n) \coloneqq n + 1 \hspace*{2em}
  \ap{decr}_p(n) \coloneqq
    \max{(0,n-1)} \hspace*{2em} 
  \ap{read}_p^r(n,\_) \coloneqq (n,n)
\end{gather*}

We verify commutativity conditions for $\ap{incr}$ and $\ap{decr}$ w.r.t.\
an initial value $n \in X_{\ctr_p}$.
By evaluating the abstract programs 
(as in Ex.~\ref{ex:nnc-commute}) 
we find that
$\ap{incr}_p$ and $\ap{incr}_p$ commute under $\top$,
$\ap{incr}_p$ and $\ap{decr}_p$ under $n>0$,
and $\ap{decr}_p$ and $\ap{decr}_p$ under $\top$.
We cannot compare $\ap{incr}$ or $\ap{decr}$ with $\ap{read}$ directly, as
they are defined in different domains. 
So we consider $\ap{incr}_p \ast \ap{id}$ and
$\ap{decr}_p \ast \ap{id}$
 in $\sf{Ctr}_p \ast \sf{Addr}_r^\N$,
while preserving soundness w.r.t.\ the concrete programs via Cor.~\ref{thm:abstract-frame-rule}. 
We can verify that, for an initial value $(n,\_) \in X_{\sf{Ctr}_p \ast \sf{Addr}_r^\N}$,
$\ap{incr}_p \ast \ap{id}$ and $\ap{read}_p^r$ never commute,
and
$\ap{decr}_p \ast \ap{id}$ and $\ap{read}_p^r$ commute under $n=0$.

Lastly, we pair $\ap{read}$ with itself,
recognizing that the two $\ap{read}$ executions should write their outputs to different locations
(this idea returns in the subsequent examples).
So with addresses $r$ and $s$,
we consider $\ap{read}_p^r \ast \ap{id}$
and $\ap{read}_p^s \ast \ap{id}$
defined in $\sf{Ctr}_p \ast \sf{Addr}_r^\N \ast \sf{Addr}_s^\N$,
and find that they commute under $\top$.

To derive concrete commutativity
from these abstract results,
we first find with symbolic execution
the soundness and capture facts
that
$\soundcapt{\sf{Ctr}_p}{\ap{incr}_p}{\tt{incr}_p}$,
$\soundcapt{\sf{Ctr}_p}{\ap{decr}_p}{\tt{decr}_p}$,
and
$\soundcapt{\sf{Ctr}_p \ast \sf{Addr}_r^\N}{\ap{read}_p^r}{\tt{read}_p^r}$
(recall the shorthand in \S\ref{sec:capture}).
Now we may immediately derive \emph{framed}
concrete commutativity conditions via Thm.~\ref{thm:commute-sound}.
For each program pair,
we characterize $\commuteconc \aA f {P^+} {\hspace*{-0.2em}g}$
(recall shorthand~\ref{def:obseq-from-abstr}):

\begin{center}
\setlength{\tabcolsep}{0.3em}
\begin{tabular}{cc|l|l}
  $f$ & $g$ & \aA & $P$ \\
  \midrule
  $\tt{incr}_p$ & $\tt{incr}_p$ & $\sf{Ctr}_p$
    & $\exists n \in \N.\ p \mapsto n$ \\
  $\tt{incr}_p$ & $\tt{decr}_p$ & $\sf{Ctr}_p$
    & $\exists n \in \N^+.\ p \mapsto n$ \\
  $\tt{decr}_p$ & $\tt{decr}_p$ & $\sf{Ctr}_p$
    & $\exists n \in \N.\ p \mapsto n$ \\
  $\tt{incr}_p$ & $\tt{read}_p^r$ & $\sf{Ctr}_p \ast \sf{Addr}_r^\N$
    & $\bot$ \\
  $\tt{decr}_p$ & $\tt{read}_p^r$ & $\sf{Ctr}_p \ast \sf{Addr}_r^\N$
    & $\exists x \in \N.\ p \mapsto 0 \ast r \mapsto x$ \\
  $\tt{read}_p^r$ & $\tt{read}_p^s$ & $\sf{Ctr}_p \ast \sf{Addr}_r^\N \ast \sf{Addr}_s^\N$
    & $\exists n,x,y \in \N.\ p \mapsto n \ast r \mapsto x \ast s \mapsto y$
\end{tabular}
\end{center}

\subsection{Two-Set}
\label{ex:two-set-progs}

Consider an unordered set
containing at most two elements of type $T \subseteq \V^*$.
We define concrete
add, remove, and member-test methods,
%
each parameterized on
the set locations $p$ and $q$
(we omit $p$ and $q$ subscripts for brevity),
as well as input/output addresses $a$ and $r$.
\begin{center}
  \input{conc_progs/set-add.tex}
  \input{conc_progs/set-rem.tex}
  \input{conc_progs/set-mem.tex}
\end{center}

\noindent
Define abstraction
$\twoset_{p,q}^T \coloneqq \abstr{
  \setbuild{S \in \pset T}
    {\abs{S} \leq 2} 
}{\pi}$ with:

\input{figures/ts-abstr.tex}
\noindent
Next we construct abstract programs $\ap{add}$ and $\ap{rem}$ in
$\sf B = \twoset_{p,q}^T \ast \sf{Addr}^T_{a}$,
and
$\ap{mem}$ in
$\sf C = \twoset_{p,q}^T \ast \sf{Addr}_a^T \ast \sf{Addr}_r^{\mathbb{B}}$:
\begin{alignat*}{2}
  \ap{add}_a(S,v) \coloneqq& \
    (S',v) \ \text{where} \
    S' = S \cup \set{v} \ \text{if} \
    \abs{S \cup \set{v}} \leq 2 \ \text{else} \ S \\
  \ap{rem}_a(S,v) \coloneqq& \
    (S \setminus \set{v}, v) \\
  \ap{mem}_a^r(S,v,\_) \coloneqq& \
  \big(S,v,
      \sf{true}\ \text{if}\
      v \in S\
      \text{else}\ \sf{false}
  \big)
\end{alignat*}

We analyze commutativity w.r.t.\
$\sf A = \twoset_{p,q}^T \ast \sf{Addr}^T_a \ast \sf{Addr}^T_r
\ast \sf{Addr}^T_b \ast \sf{Addr}^T_s$
(though only the $\ap{mem}$/$\ap{mem}$ pair utilizes all four \sf{Addr} domains).

{

\newcommand*{\setcomm}[4]
{\ensuremath{\commuteabs{\sf A}{\ap{#1} #2}{Q}{\ap{#3} #4}}}

\begin{center}
\begin{tabular}{rl|l}
  \multicolumn{2}{c|}{Programs} &
  Condition on $(S,u,\_,v,\_) \in X_\sf A$ \\
  \midrule
  $\ap{add}_a$ & $\ap{add}_b$ &
    $S = \emptyset \vee \abs{S} = 2 \vee
     u \in S \vee v \in S \vee u = v$ \\
  $\ap{add}_a$ & $\ap{rem}_b$ &
    \makecell[tl]{
    $\abs{S} < 2 \wedge u \neq v\ \vee$ \\
     $\abs{S} = 2 \wedge (u,v \notin S \vee u \in S \wedge u \neq v)$} \\
  $\ap{rem}_a$ & $\ap{rem}_b$ & $\top$ \\
  $\ap{mem}_a^r$ & $\ap{add}_b$ &
    $u \neq v \vee v \in S \vee \abs{S} = 2$ \\
  $\ap{mem}_a^r$ &$\ap{rem}_b$ &
    $u \neq v \vee v \notin S$ \\
  $\ap{mem}_a^r$ & $\ap{mem}_b^s$ & $\top$ \\
\end{tabular}
\end{center}

\noindent
After proving
that
$\soundcapt{\sf B}{\ap{add}_a}{\tt{add}_a}$,
$\soundcapt{\sf B}{\ap{rem}_a}{\tt{rem}_a}$,
and
$\soundcapt{\sf C}{\ap{mem}_a^r}{\tt{mem}_a^r}$,
we again can
derive concrete commutativity conditions
for \tt{add}, \tt{rem}, and \tt{mem}.
In particular, we
use the observational equivalence
induced by \aA\ as our concrete eq.\ relation,
which ensures that
concrete two-sets with the same values in different orders
are considered equivalent.
We omit for brevity the concrete conditions,
which each amount to
``the program pair commutes w.r.t.\ $[\purview\aA,\sim_\aA]$ on $h$
if $\pi(h)$ satisfies the corresponding abstract condition.''

}

\subsection{Linked Stack}
\label{subsec:full-ex-stk}

We now expand upon the example from the overview (\S\ref{sec:overview})
of a stack of non-\vnull\ values,
implemented as a linked list accessible at address $\ell$.
We consider its \code{push}, \code{pop}, and \code{peek} methods,
where \code{pop} fails on an empty stack.
To account for the lack of an \code{is\_empty} operation,
\code{peek} outputs \vnull\ on an empty stack.

\begin{center}
  \input{conc_progs/stk-push.tex}
  \input{conc_progs/stk-pop.tex}
  \input{conc_progs/stk-peek.tex}
\end{center}

Once again, we abstraction define
$\sf{Stk}_\ell \coloneqq \abstr{\mathrm{list}(\V^*)}{\pi}$ where
\begin{align*}
  \pi(h) \coloneqq\ & s\ \text{if}\ \PsiStack^+(s,h)\ \text{for}\
  s \in \mathrm{list}(\V^*)\ \text{else}\ \opurv \\[4pt]
  \PsiStack(s) \coloneqq\ &
    \exists p \in \L + \vnull.\
    \ell \mapsto p \ast
    \psi(s,p) \\
  \psi(v \cons s, p) \coloneqq\ &
    \exists q \in \L + \vnull.\
    p \neq \vnull \ast
    p \mapsto (v,q) \ast
    \psi(s,q) \\
  \psi([],p) \coloneqq\ &
    p = \vnull
\end{align*}

\noindent
Define
abstract $\ap{push}$/$\ap{pop}$ in
$\sf B \coloneqq \sf{Stk}_\ell \ast \sf{Addr}_a^{\V^*}$\hspace*{-0.2em},
and $\ap{peek}$ in $\sf C \coloneqq  \sf{Stk}_\ell \ast \sf{Addr}_a^{\V}$:
\begin{alignat*}{4}
  \ap{push}_\ell^a(s,v) \coloneqq&\ (v \cons s, v) &
  \ap{pop}_\ell^a([],\_) \coloneqq&\ \bot & \hspace*{1em} 
  \ap{peek}_\ell^a(v \cons s, \_) \coloneqq&\ (v \cons s, v) \\
  && \hspace*{1em} \ap{pop}_\ell^a(v \cons s, \_) \coloneqq&\ (s, v) &
  \ap{peek}_\ell^a([],\_) \coloneqq&\ ([], \vnull)
\end{alignat*}

\noindent
Next we verify commutativity conditions for each abstract program pair w.r.t.\
$\sf A = \sf{Stk}_\ell \ast \sf{Addr}^U_a
\ast \sf{Addr}^V_b$,
where $U$ and $V$ depend on the pair.
\begin{center}
\setlength\tabcolsep{3pt}
\begin{tabular}{rl|lll}
  \multicolumn{2}{c|}{Programs} & $U$ & $V$ &
  Cond.\ on $(s,u,v) \in X_\sf A$ \\
  \midrule
  $\ap{push}_\ell^a$ & $\ap{push}_\ell^b$
    & $\V^*$ & $\V^*$ &
    $u = v$ \\
  $\ap{push}_\ell^a$ & $\ap{pop}_\ell^b$
    & $\V^*$ & $\V^*$ &
    $s = u \cons \_$ \\
  $\ap{push}_\ell^a$ & $\ap{peek}_\ell^b$
    & $\V^*$ & $\V$ &
    $s = u \cons \_$ \\
  $\ap{pop}_\ell^a$ & $\ap{pop}_\ell^b$
    & $\V^*$ & $\V^*$ &
    $\exists x.\ s = x \cons x \cons \_$ \\
  $\ap{pop}_\ell^a$ &$\ap{peek}_\ell^b$
    & $\V^*$ & $\V$ &
    $\exists x.\ s = x \cons x \cons \_$ \\
  $\ap{peek}_\ell^a$ & $\ap{peek}_\ell^b$
    & $\V$ & $\V$ &
    $\top$ \\
\end{tabular}
\end{center}

\noindent
We then find through
symbolic execution that
$\soundcapt{\sf B}{\ap{push}_\ell^a}{\tt{push}_\ell^a}$,
$\soundcapt{\sf B}{\ap{pop}_\ell^a}{\tt{pop}_\ell^a}$,
and $\soundcapt{\sf C}{\ap{peek}_\ell^a}{\tt{peek}_\ell^a}$.
Finally, we freely derive commutativity conditions
for each concrete pair
w.r.t.\ $[\purview{\aA},\sim_\aA]$,
which we again omit for brevity.

\subsection{Composition of Stack and Counter}
\label{subsec:full-ex-comp}

Suppose we augment our stack data structure
with a counter to track the stack's size.
The concrete operations for this structure would be
{\small
\begin{align*}
  \tt{cpush}_{\ell,p}^a \coloneqq\ & \tt{push}_\ell^a \mathop{;} \tt{incr}_p
  \hspace*{1em}
  & \tt{cpeek}_\ell^a \coloneqq\ & \tt{peek}_\ell^a
  \\
  \tt{cpop}_{\ell,p}^a \coloneqq\ & \tt{pop}_\ell^a \mathop{;} \tt{decr}_p
  & \tt{csize}_p^r \coloneqq\ & \tt{read}_p^r
\end{align*}
}

\noindent
We can similarly define abstract programs with composition.
Namely, the following are defined in 
$\sf{Stk}_\ell \ast \ctr_p$ (with $\sf{Addr}$
domains joined on as appropriate):
\begin{align*}
  \ap{cpush}_{\ell,p}^a \coloneqq\ & \ap{push}_\ell^a \ast \ap{incr}_p
  \hspace*{1em}
  & \ap{cpeek}_\ell^a \coloneqq\ & \ap{peek}_\ell^a \ast \ap{id}
  \\
  \ap{cpop}_{\ell,p}^a \coloneqq\ & \ap{pop}_\ell^a \ast \ap{decr}_p
  & \ap{csize}_p^r \coloneqq\ & \ap{id} \ast \ap{read}_p^r
\end{align*}

We analyze commutativity of these abstract programs in
$\sf A = \sf{Stk}_\ell \ast \ctr_p \ast \sf{Addr}^U_a \ast \sf{Addr}^V_b$.
For each pair, we leverage Lem.~\ref{lem:commute-conj}
to derive a condition for free from our prior analyses
($\ell$ and $p$ subscripts are omitted):

\setlength\tabcolsep{3pt}
\begin{center}
\begin{tabular}{rl|lll}
  \multicolumn{2}{c|}{Programs} & $U$ & $V$ &
  Cond.\ on $(s,n,u,v) \in X_\sf A$ \\
  \midrule
  $\ap{cpush}^a$ & $\ap{cpush}^b$
    & $\V^*$ & $\V^*$ &
    $u = v$ \\
  $\ap{cpush}^a$ & $\ap{cpop}^b$
    & $\V^*$ & $\V^*$ &
    $s = u \cons \_ \wedge n > 0$ \\
  $\ap{cpush}^a$ & $\ap{cpeek}^b$
    & $\V^*$ & $\V$ &
    $s = u \cons \_$ \\
  $\ap{cpush}^a$ & $\ap{csize}^b$
    & $\V^*$ & $\N$ &
    $\bot$ \\
  $\ap{cpop}^a$ & $\ap{cpop}^b$
    & $\V^*$ & $\V^*$ &
    $\exists x.\ s = x \cons x \cons \_$ \\
  $\ap{cpop}^a$ &$\ap{cpeek}^b$
    & $\V^*$ & $\V$ &
    $\exists x.\ s = x \cons x \cons \_$ \\
  $\ap{cpop}^a$ &$\ap{csize}^b$
    & $\V^*$ & $\N$ &
    $s = \_ \cons \_ \wedge n = 0$ \\
  $\ap{cpeek}^a$ & $\ap{cpeek}^b$
    & $\V$ & $\V$ &
    $\top$ \\
  $\ap{cpeek}^a$ & $\ap{csize}^b$
    & $\V$ & $\N$ &
    $\top$ \\
  $\ap{csize}^a$ & $\ap{csize}^b$
    & $\N$ & $\N$ &
    $\top$ \\
\end{tabular}
\end{center}

\noindent
For our requisite soundness and capture facts,
we use compound soundness (Thm.~\ref{thm:conjoin-soundness})
to derive for free that
$\soundcapt
  {\sf A}
  {\ap{cpush}_{\ell,p}^a}
  {\tt{cpush}_{\ell,p}^a}$,
$\soundcapt
  {\sf A}
  {\ap{cpop}_{\ell,p}^a}
  {\tt{cpop}_{\ell,p}^a}$,
$\soundcapt
  {\sf A}
  {\ap{cpeek}_{\ell}^a}
  {\tt{cpeek}_{\ell}^a}$,
and
$\soundcapt
  {\sf A}
  {\ap{csize}_{p}^a}
  {\tt{csize}_{p}^a}$.
With this and our abstract commutativity conditions,
we may freely derive concrete conditions
for \tt{cpush}, \tt{cpop}, \tt{cpeek}, and \tt{csize}.
One interesting case is the
$\ap{cpop}$/$\ap{csize}$ condition,
which demands a nonempty stack with size 0.
This cannot occur with a correctly initialized stack and counter,
meaning in practice
$\tt{cpop}$ and $\tt{csize}$ never commute.

%% file: conc_progs/ctr-incr.tex
\begin{minipage}[t]{1.2in}%
  {\small\vspace{-\baselineskip-\abovedisplayskip}
  \begin{alignat*}{2}
    & \tt{incr}_p \coloneqq \\[-3pt]
    & \langInd \langLet\ c = \langDeref p\ \langIn \\[-3pt]
    & \langInd \langLet\ i = c + 1\ \langIn \\[-3pt]
    & \langInd p \langWrite i
  \end{alignat*}}
\end{minipage}

%% file: conc_progs/ctr-decr.tex
\begin{minipage}[t]{1.5in}%
  {\small\vspace{-\baselineskip-\abovedisplayskip}
  \begin{alignat*}{2}
    & \tt{decr}_p \coloneqq 
    \langLet\ c = \langDeref p\ \langIn \\[-3pt]
    & \langInd \langLet\ i = c - 1\ \langIn \\[-3pt]
    & \langInd \langIf\ i < 0\ \langThen\ \langSkip \\[-3pt]
    & \langInd \langElse\ p \langWrite i
  \end{alignat*}}
\end{minipage}

%% file: conc_progs/ctr-read.tex
\begin{minipage}[t]{1.2in}%
  {\small\vspace{-\baselineskip-\abovedisplayskip}
  \begin{alignat*}{2}
    & \tt{read}_p^r \coloneqq \\[-2pt]
    & \langInd \langLet\ c = \langDeref p\ \langIn \\[-3pt]
    & \langInd r \langWrite c
  \end{alignat*}}
\end{minipage}

%% file: conc_progs/set-add.tex
\begin{minipage}[t]{1.5in}%
  {\small\vspace{-\baselineskip-\abovedisplayskip}
  \begin{alignat*}{2}
    & \tt{add}_a \coloneqq 
    \langLet\ u = \langDeref p\ \langIn \\[-3pt]
    & \langInd \langLet\ v = \langDeref q\ \langIn \\[-3pt]
    & \langInd \langLet\ x = \langDeref a\ \langIn \\[-3pt]
    & \langInd \langIf\ u = x \vee v = x\ \langThen \\[-3pt]
    & \langInd \langInd \langSkip\ \langElse \\[-3pt]
    & \langInd \langIf\ u = \vnull\ \langThen \\[-3pt]
    & \langInd \langInd p \langWrite x\ \langElse \\[-3pt]
    & \langInd \langIf\ v = \vnull\ \langThen \\[-0.4em]
    & \langInd \langInd q \langWrite x\ \langElse \\[-3pt]
    & \langInd \langSkip
  \end{alignat*}}
\end{minipage}

%% file: conc_progs/set-rem.tex
\begin{minipage}[t]{1.5in}%
  {\small\vspace{-\baselineskip-\abovedisplayskip}
  \begin{alignat*}{2}
    & \tt{rem}_a \coloneqq 
     \langLet\ u = \langDeref p\ \langIn \\[-3pt]
    & \langInd \langLet\ v = \langDeref q\ \langIn \\[-3pt]
    & \langInd \langLet\ x = \langDeref a\ \langIn \\[-3pt]
    & \langInd \langIf\ u = x\ \langThen \\[-3pt]
    & \langInd \langInd p \langWrite \vnull\ \langElse \\[-3pt]
    & \langInd \langIf\ v = x\ \langThen \\[-0.4em]
    & \langInd \langInd q \langWrite \vnull\ \langElse \\[-3pt]
    & \langInd \langSkip
  \end{alignat*}}
\end{minipage}

%% file: conc_progs/set-mem.tex
\begin{minipage}[t]{1.5in}%
  {\small\vspace{-\baselineskip-\abovedisplayskip}
  \begin{alignat*}{2}
    & \tt{mem}_a^r \coloneqq 
    \langLet\ u = \langDeref p\ \langIn \\[-3pt]
    & \langInd \langLet\ v = \langDeref q\ \langIn \\[-3pt]
    & \langInd \langLet\ x = \langDeref a\ \langIn \\[-3pt]
    & \langInd \langIf\ u = x\ \langThen \\[-3pt]
    & \langInd \langInd r \langWrite \langTrue\ \langElse \\[-3pt]
    & \langInd \langIf\ v = x\ \langThen \\[-3pt]
    & \langInd \langInd r \langWrite \langTrue\ \langElse \\[-3pt]
    & \langInd r \langWrite \langFalse
  \end{alignat*}}
\end{minipage}

%% file: conc_progs/stk-push.tex
\begin{minipage}[t]{1.5in}%
  {\small\vspace{-\baselineskip-\abovedisplayskip}
  \begin{alignat*}{2}
    & \tt{push}_\ell^a \coloneqq 
     \langLet\ p = \langDeref \ell\ \langIn \\[-3pt]
    & \langInd \langLet\ v = \langDeref a\ \langIn \\[-3pt]
    & \langInd \langLet\ x = (v,p)\ \langIn \\[-3pt]
    & \langInd \langLet\ q = \langRef\ x\ \langIn \\[-3pt]
    & \langInd \ell \langWrite q
  \end{alignat*}}
\end{minipage}

%% file: conc_progs/stk-pop.tex
\begin{minipage}[t]{1.5in}%
  {\small\vspace{-\baselineskip-\abovedisplayskip}
  \begin{alignat*}{2}
  & \tt{pop}_\ell^a \coloneqq 
   \langLet\ p = \langDeref \ell\ \langIn \\[-3pt]
  & \langInd \langIf\ p = \vnull\ \langThen \\[-3pt]
  & \langInd \langInd \langFail\ \langElse \\[-3pt]
  & \langInd \langLet\ x = \langDeref p\ \langIn \\[-3pt]
  & \langInd \langLet\ v = \langFst\ x\ \langIn \\[-3pt]
  & \langInd \langLet\ q = \langSnd\ x\ \langIn \\[-3pt]
  & \langInd \ell \langWrite q\ \langSeq\
    \langFree\ p\ \langSeq \\[-3pt]
  & \langInd a \langWrite v
  \end{alignat*}}
\end{minipage}

%% file: conc_progs/stk-peek.tex
\begin{minipage}[t]{1.5in}%
  {\small\vspace{-\baselineskip-\abovedisplayskip}
  \begin{alignat*}{2}
  & \tt{peek}_\ell^a \coloneqq 
   \langLet\ p = \langDeref \ell\ \langIn \\[-3pt]
  & \langInd \langIf\ p = \vnull\ \langThen \\[-3pt]
  & \langInd \langInd a \langWrite \vnull\ \langElse \\[-3pt]
  & \langInd \langLet\ x = \langDeref p\ \langIn \\[-3pt]
  & \langInd \langLet\ v = \langFst\ x\ \langIn \\[-3pt]
  & \langInd a \langWrite v
  \end{alignat*}}
\end{minipage}

%% file: sections/related-work.tex
\section{Related Work}

To the best of our knowledge there are no existing works 
on commutativity reasoning specifically geared 
toward heap-based programs,
aside from preliminary work by Pincus~\cite{pincus-thesis} which is limited to deterministic programs.
We discuss some works in \S\ref{sec:introduction};
and here discuss those and others in more detail.

\emph{Commutativity without the heap.}
Some prior works focused on verifying and even inferring 
commutativity properties, though without a heap memory model. 
These include aforementioned work such as 
Kim and Rinard~\cite{rinard_linked_structures} who 
verified commutativity properties in two steps:
verifying an implementation satisfies its ADT specs, and then 
verifying commutativity of the ADT spec. 
Further in that direction, Bansal \emph{et al.}~\cite{bansal_tacas,JAR20}
showed that commutativity conditions of ADT specs could be 
synthesized, and introduced the tool {\sc Servois},
which was later improved as {\sc Servois2}~\cite{ATVA23}.
Chen \emph{et al.}~\cite{veracity_paper} adapted these approaches
to perform commutativity analysis on (non-heap) imperative programs for parallelization.
%
Koskinen and Bansal~\cite{VMCAI2021} also verified commutativity,
but not of heap-based programs.

\emph{Commutativity with the heap.}
P\^{i}rlea \emph{et al.}~\cite{sergey_sharding}
describe the {\sc CoSplit} tool, which performs a 
commutativity analysis on a smart contract language. 
Their analysis determines commutativity of heap operations,
but only when commutativity can be determined based on heap 
ownership. Thus, their approach could not handle 
the examples in this paper including the non-negative counter, which 
involves commutative updates to the same heap cell.
Eilers \emph{et al.}~\cite{EilersPLDI2023} 
leverage commutativity modulo user-specified abstractions to
prove information flow security in the space of relational concurrent separation logic.

\emph{Exploiting commutativity for verification.}
Commutativity is a widely used abstraction in many 
verification tools, which use commutativity specifications 
as user-provided inputs. 
In the context of concurrent programs, 
QED~\cite{qed} and later CIVL~\cite{civl} both 
use commutativity (more specifically, left/right movers) to build atomic sections.
The {\sc Anchor}~\cite{anchor} verifier also uses commutativity for a more automated 
approach of verifying concurrent programs.
Commutativity is also used for proofs of parameterized programs \cite{FarzanPOPL2024},
operational-style proofs of concurrent objects~\cite{quotients}
and termination of concurrent programs \cite{FarzanCAV2023}. 
Abstract commutativity relations have also been used in reductions for verification~\cite{FarzanPOPL2023}. 
A summary of the use of commutativity in verification was given by~\cite{FarzanLICS2023}.

\emph{Abstract domains for the heap.}
Other works have focused on the intersection of 
abstract interpretation and heap logics, although 
they do not specifically target commutativity,
and generally seek to abstract
the semantics of separation logic
rather than reasoning about particular allocated structures.
Sims \cite{Sims06} constructs a detailed
abstract domain for representing 
separation logic heap predicates.
Calcagno \emph{et al.}~\cite{abstract-sep-logic} introduce separation algebras,
which generalize the semantics of separation logic.

%% file: sections/conclusion.tex
\section{Conclusion}

We have demonstrated how
commutativity analysis on concrete heap programs
can be reduced to much simpler reasoning
about mathematical objects
in an abstract space.
We have designed our abstract domain to
account for allocation nondeterminism,
concrete observational equivalence,
and improperly allocated heap structures.
We formalized this domain,
laid out the abstract semantics
and program soundness relation,
established a sound commutativity theorem,
and introduced composition of abstractions
for multi-structure program settings.
We then worked through several examples illustrating the
convenience of analyzing abstract program commutativity
and deriving concrete results.
Our work has been implemented in Coq 
and is available in the 
\VMCAI{supplement.}%
\ARXIV{\href{\supplementlink}{supplement}.}%

In future work we plan to pursue automation by implementing an abstract interpreter, and 
exploring how it can be combined with existing commutativity synthesis 
techniques in the abstract domain~\cite{bansal_tacas,ATVA23}.
Furthermore, we will explore how our model of data structure abstraction
could be applicable to various kinds of heap-style reasoning
other than commutativity.
We are also interested in augmenting our abstract domain
to feature an abstract value subset lattice;
this would admit nondeterministic abstract programs,
and enable reasoning about bounded divergence commutativity.

%% file: main/acknowledgements.tex
We thank Marco Gaboardi, David Naumann,
VFC,
and the anonymous reviewers for their feedback on earlier versions of this draft.
Both authors were partially supported by NSF award \#2008633.
Koskinen was partially supported by NSF award \#2315363.
Pincus was partially supported by NSF award \#1801564.

%% file: main/disclosures.tex
The authors have no competing interests to declare. 

%% file: appendices/isomorphism.tex
\section*{Appendix}
\label{appdx:isomorphism}

To share a specific example of isomorphism,
as well as an alternative way to construct abstract domains,
we return to our two-set example (Ex.~\ref{ex:two-set}),
again containing values of type $T \subseteq \V^*$
at addresses $p$ and $q$.
Defining $\pi$ in that example involved explicitly tracking
different versions of the structure,
taking care that the concrete order of values
did not affect the abstract mapping.

Here, we take a different, perhaps more intuitive approach.
We will explicitly construct an equivalence relation on heaps,
and derive $X$ and $\pi$ from there.
Here is said equivalence relation,
using a heap predicate similar to what Ex.~\ref{ex:two-set} used:
\begin{align*}
  h \sim h' \coloneqq\ &
    \exists u, v.\
    \wedge
    \begin{cases}
      \PsiSetEq^+\big((u,v),h\big) \\
      \PsiSetEq^+\big((u,v),h'\big) \vee
        \PsiSetEq^+\big((v,u),h'\big)
    \end{cases} \\
  \PsiSetEq\big((u,v),\cdot\big) \coloneqq\ &
      p \mapsto u \ast q \mapsto v \ast
      (u,v \in T + \vnull) 
    \ast
    (u = v \implies u = \vnull)
\end{align*}
Our abstract values will be the equivalence classes of heaps containing two-sets,
and the projection function will extend the canonical projection.
Specifically, $\sf{Setq}^T_{p,q} = \abstr{\H(\PsiSetEq^+)/{\sim}}{\pi}$,
where $\pi(h) = [h]_\sim$ if $\PsiSetEq^+(h)$, else $\opurv$.

Indeed, we show in Coq that $\sf{Set}^T_{p,q} \cong \sf{Setq}^T_{p,q}$.
Comparing the two constructions,
we see how this eq.\ relation approach
implicitly and unambiguously captures 
each possible set cardinality,
as well as concrete order irrelevance.
On the other hand, explicit representations of abstract values are often desirable for human comprehension
(e.g.\ sets of size 0--2),
but 
equivalence classes are more opaque in their meaning.
This complicates the construction and analysis of abstract programs.
For this reason, we only use $\twoset_{p,q}^T$
when working with two-set programs in 
\S\ref{ex:two-set-progs}.